\documentclass[aps, prl, groupaddress, twocolumn
]{revtex4-1}
\usepackage{graphicx}
\usepackage{dcolumn}
\usepackage{bm}
\usepackage{hyperref}
\usepackage[normalem]{ulem}
\usepackage{color}
\usepackage{amsmath}
\usepackage{amsfonts}
\usepackage{amssymb}%

\begin{document}

\title{Gauge invariance and the interpretation of
inter- and intraband processes \\in
high-order harmonic generation from bulk solids}

\author{P\'{e}ter F\"{o}ldi}
\affiliation{Department of Theoretical Physics, University of Szeged,\\ Tisza Lajos k\"{o}r%
\'{u}t 84-86, H-6720 Szeged, Hungary}
\affiliation{ELI-ALPS, ELI-HU Non-profit Ltd., Dugonics t\'{e}r 13, H-6720 Szeged, Hungary}

%\affiliation{Department of Theoretical Physics, University of Szeged, Tisza Lajos k\"{o}r\'{u}t 84, H-6720 Szeged, Hungary}
%\affiliation{ELI-ALPS, ELI-HU Non-profit Ltd., Dugonics t\'{e}r 13, H-6720 Szeged, Hungary}%	

%\pacs{42.50.Gy, 42.65.Ky}

\begin{abstract}
A theoretical model for high-order harmonic generation (HHG) in bulk solids is considered. Our approach treats laser-induced inter- and intraband currents on an equal footing. The sum of these currents is the source of the high-order harmonic radiation, and does not depend on the particular electromagnetic gauge we choose to describe the process. On the other hand, as it is shown using analytic and numerical calculations, the distinction between intra- and interband dynamics is gauge dependent, implying that the interpretation of the process of HHG using these terms requires carefulness.
\end{abstract}

\maketitle

%\section{Introduction}

\label{introsec}
\emph{Introduction }  The idea that solid state targets illuminated by strong, short near infrared pulses can produce coherent X-ray radiation \cite{KB95} or even attosecond electromagnetic bursts
\cite{FK96} preceded the first experiments that demonstrated the appearance of high-order harmonics (up to order of 25) in the spectra of strongly driven solid state samples. In Ref.~\cite{GDiCSADiMR11} a wide bandgap ZnO target (3.2 eV) was excited by a pulse with central wavelength of 3.25 $\mu$m,  meaning that at least 9 photons are required for an excitation from the valance to the conduction band [see also \cite{GNDiCSSADiMR14} for more details].
Recently, a direct comparison of high-order harmonic generation in the solid and gas phases of argon and krypton has been reported \cite{NGWBScGR16}.

Theoretically, semiconductor Bloch-equations were applied to describe the problem \cite{GMK08}, a closed-form expression were given to the subcycle-resolved transition rate of electrons between bands \cite{HaIv13}, appearance of attosecond pulses were predicted \cite{HSH14}, semiclassical \cite{VMcOKCB14} and a saddle-point \cite{VMcOCB15} analysis were performed and the role of an indirect bandgap was also investigated \cite{YaNo16}.

\bigskip

The intensity of the exciting laser pulse was below the damage threshold in Ref.~\cite{GDiCSADiMR11}. Although high-order harmonic generation
(HHG) is known to be possible also in plasmas emerging from solid state surfaces as a consequence of intense electromagnetic radiation [see e.g.,~\cite{TG09} for a review],
here we focus on the case when
a system of electrons in a static, periodic potential
interacts with the exciting laser field \cite{HIY15}.

Neglecting relaxation and multiparticle effects, the response of the
electronic subsystem in a solid to the external laser pulse is usually
categorized as follows:
(1) Interband transitions,
(2) Laser-driven intraband motion,
(3) Dynamical Bloch-oscillations.
The first phenomenon is the usual optical generation
of charge carriers, but this is already a nonlinear effect in the range of
HHG. Points 2 and 3 above are closely
related, it is only the strength of the laser field that determines whether
dynamical Bloch oscillations appear or not. As a quasiclassical picture, the external field of the laser changes the crystal momentum
of the electrons. More precisely, as described by the acceleration theorem
\cite{K05,AH11}, the change of the crystal momentum can be written as
\begin{equation}
\hbar\frac{d\mathbf{k}(t)}{dt}=e_{0}\mathbf{F}(t), \label{accelerationeq}
\end{equation}
where $\mathbf{F}(t)$ represents the time dependent external
electric field, and $e_{0}$ denotes the charge of the electron. (That is,
here, and in the following, $e_{0}$ is negative.) When the amplitude of this
oscillatory motion is large enough to cross the boundary of the first
Brillouin zone, Bragg-reflection occurs, which can
be termed as dynamical Bloch-oscillation \cite{M11,FBY13}.
It is important to note that mechanisms 1-3 are coupled, practically none of
them appears on its own \cite{HSH14}.

\bigskip

As we show in the
following, although separating inter- and intraband dynamics is very useful
for the intuitive interpretation of the physical processes, it depends on the
choice of the electromagnetic gauge. In other words, gauge transformations
(which, obviously, do not change physically relevant results) mix the inter-
and intraband dynamics.
%%%%%%%%%%%%%%%%%%%%%%%%%%%%%%%%%%%%%%%%%%%%%%%%%%%%%%%%%%%%%%%%%%%%%%%%%%%%%%%%%%%%%%%%%%%%

\bigskip

The Hamiltonian describing the electron in a periodic
potential (representing the solid) and interacting with an external field can
be written as:
\begin{align}
H(t)=&H_{0}+H_{ext}(t)=\frac{1}{2m}(\mathbf{p-}e_{0}\mathbf{A})^{2}%
+V(\mathbf{r})+e_{0}\Phi\nonumber \\
=&\frac{\mathbf{p}^{2}}{2m}+V(\mathbf{r})+H_{ext}%
(t),\label{Ham}%
\end{align}
where $V(\mathbf{r})$ is the lattice-periodic potential of the crystal,
while $H_{ext}(t)$ takes the interaction with the external field into account, via
the electromagnetic potentials. Note the appearance of the kinetic momentum $\mathbf{p}_{kin}=m\mathbf{v=p-}%
e_{0}\mathbf{A}$ in Eq.~(\ref{Ham}), which is seen to be different from
the canonical $\mathbf{p}=-i\hbar\nabla$\, depending on the gauge.

The eigenstates of the field-free Hamiltonian obeying $H_{0}|n,\mathbf{k}%
\rangle=E_{n}(\mathbf{k})|n,\mathbf{k}\rangle,$ in coordinate representation, take the form $\Psi
_{n,\mathbf{k}}(\mathbf{r})=\exp(i\mathbf{kr})u_{n,\mathbf{k}}(\mathbf{r}%
)/\sqrt{\mathcal{V}},$ where $n$ denotes the band
index of these Bloch-states, $u_{n,\mathbf{k}%
}(\mathbf{r})$ are lattice periodic functions and $\mathcal{V}$ is the crystal
volume. We assume here, as usual, periodic boundary conditions, and then the
$\mathbf{k}$-space will not be continuous, we obtain a discrete, densely
spaced series of $\mathbf{k}$ vectors. The $\mathbf{k}$-dependent
eigenenergies $E_{n}(\mathbf{k})$ provide the dispersion relations for the
various bands.

In order to specify the interaction term $H_{ext}(t)$, we have to choose a
gauge. Clearly, physical predictions based on exact solutions without
approximations must be the same in all gauges. We assume here that the spatial
variation of the external field $\mathbf{F(}t\mathbf{)}$ can be neglected,
since the wavelengths of the exciting infrared pulses are much longer than the
lattice constants in a crystal. Then one of the appropriate gauges is the
velocity ($v$) gauge, where the vector and scalar potentials of the field
$\mathbf{F}(t)$ are $\mathbf{A}^{v}(t)=-\int_{-\infty}^{t}\mathbf{F}%
(t^{\prime})dt^{\prime},\Phi^{v}=0$. With this choice, we have
\begin{equation}
H_{ext}^{v}(t)=\frac{1}{2m}\left[  -2e_{0}\mathbf{p}\mathbf{A}^{v}%
\mathbf{(}t\mathbf{)}+e_{0}^{2}A^{v2}(t)\right]  .\label{Hv}%
\end{equation}
The second widely used option is the length ($l$) gauge, in which
$\mathbf{A}^{l}=0,\Phi^{l}=-\mathbf{rF}(t)$, obtained from the $v$
gauge by using the gauge function $\Lambda=-\mathbf{rA}%
^{v}(t)$. This yields
\begin{equation}
H_{ext}^{l}(t)=-e_{0}\mathbf{F}(t)\mathbf{r}.\label{Hl}%
\end{equation}
(Note that both of these choices belong to the class of Coulomb gauges.) From
now on, we shall use the notation $\mathbf{A}(t)\equiv\mathbf{A}^{v}(t)$.
It should be kept in mind that electron wavefunctions
in these gauges are connected by a space-time dependent unitary transfomation:
\begin{equation}
\Psi^{l}(\mathbf{r,t})=U(\mathbf{r},t)\Psi^{v}(\mathbf{r},t)=\exp\left[
-\frac{i}{\hbar}e_{0}\mathbf{A(}t\mathbf{)}\mathbf{r}\right]  \Psi
^{v}(\mathbf{r},t).\label{lv}%
\end{equation}
In the following we
will consider pulsed excitation of duration $T,$ and assume that both
$\mathbf{F}(t)$ and $\mathbf{A}(t)$ vanish outside the interval of $[0,T].$
This means that states in the two gauges coincide for $t<0$ and $t>T$ [when
$U(\mathbf{r},t)$ is the identity]. Note that this choice avoids gauge-related
ambiguities for $t>T$ \cite{M02,RB04,YM10}.

Under the effect of $H_{ext}^{g},$ any initial state will evolve in time, and
can be expanded as a time dependent linear combination of the unperturbed
eigenstates:
As $H_{ext}^{g}$ is gauge dependent, the expansion coefficients as well as the
states $\left\vert \Psi^{g}(t)\right\rangle $ shall depend on the gauge
chosen.
We use the following
notations in the two gauges of interest here:
\begin{equation}
\left\vert \Psi^{v}(t)\right\rangle =\sum\limits_{n\mathbf{k}}c_{n\mathbf{k}%
}(t)|n,\mathbf{k}\rangle, \  \left\vert \Psi^{l}(t)\right\rangle
=\sum\limits_{n\mathbf{k}}b_{n\mathbf{k}}(t)|n,\mathbf{k}\rangle
.\label{Psiexp}%
\end{equation}
The following (invertible) relation holds
between the expansion coefficients:%
\begin{equation}
b_{n^{\prime}\mathbf{k}^{\prime}}(t)=\sum\limits_{n\mathbf{k}%
}\left\langle n^{\prime}\mathbf{k}^{\prime}\right\vert \exp\left[
-ie_{0}\mathbf{A(}t\mathbf{)}\mathbf{r/}\hbar\right]  |n,\mathbf{k}\rangle
c_{n\mathbf{k}}(t).
\label{coefflv}%
\end{equation}

Before the arrival of the exciting laser pulse ($t<0$), the solid state target can be assumed to be in thermal equilibrium. This initial condition cannot be described by a pure quantum mechanical state, we have to consider a (single particle) density operator, which is diagonal in the eigenstates of the unperturbed Hamiltonian  $H_0:$
\begin{equation}
\rho(t=0)=\sum_{n\mathbf{k}} |n,\mathbf{k}\rangle \langle n,\mathbf{k}| f[E_{n}(\mathbf{k})],
\label{rho0}
\end{equation}
where the relative statistical weights of the projectors is determined by the Fermi function $f$.  For wide-bandgap target materials at room temperature, $f[E_{n}(\mathbf{k})]$ is practically unity for the valence band, and zero for the conduction bands. That is, to a very good approximation, we can write
$\rho(t=0)=  \sum_{\mathbf{k}}  |n_0,\mathbf{k}\rangle \langle n_0,\mathbf{k}|,$
where $n_0$ corresponds to the valence band.
Similarly to the pure quantum mechanical states, the density operator is also gauge dependent, the analogue of Eq.~(\ref{lv}) reads
$\rho^l(t) = U(t) \rho^v(t) U^\dagger(t)$
and the time evolution of the density operator is given by the von-Neumann equation
$i\hbar\frac{\partial}{\partial t}\rho^{g}(t)=\left[H^{g}(t), \rho^{g}(t) \right].$
For the sake of simplicity, let us use the velocity gauge, when (in dipole approximation) the $A^2$ term in the Hamiltonian commutes with $\rho^{v}$,  and the remaining part does not mix states with different indices $\mathbf{k}$ (see the Appendix for more details). Therefore we have:
\begin{equation}
\rho^{v}(t)=\sum_{nn'\mathbf{k}} c_{n\mathbf{k}}(t) c_{n^{\prime}\mathbf{k}}^{*}(t) |n,\mathbf{k}\rangle \langle n',\mathbf{k}|,
\label{rhot}
\end{equation}
where $\sum_n |c_{n\mathbf{k}}(t)|^2=1$, and the time dependence of these coefficients is determined by the von-Neumann equation [with the initial conditions of $c_{n\mathbf{k}}(0)=\delta_{n,n_0}$]. Since the trace of the density operator is preserved during the dynamics, our choice of normalization means that $\mathrm{Tr}\rho$ equals the number $\mathcal{N}$ of valence band states in the first Brillouin zone.
Note that this approach does not provide absolute field strengths, exact "number of photons" in the HHG modes (unlike Ref.~\cite{GCVF16}), it is only the relative weight of the high harmonics that can be obtained. That is the reason why we are free to choose the normalization condition for the initial density matrix.

\bigskip

%%%%%%%%%%%%%%%%%%%%%%%%%%%%%%%%%%%%%%%%%%%%%%%%%%%%%%%%%%%%%%%%%%%%%%%%%%%%%%%%%%%%%%%%%%%%%%

\emph{High-order harmonic generation }
Moving charges mean the source of the high harmonic radiation, in other words it is the
expectation value of the time dependent current (density) that is to be calculated:
\begin{equation}
\mathbf{J:=}\frac{e_{0}}{\mathcal{V}m}\mathrm{Tr}\left[ \rho^g \mathbf{p}%
_{kin}\right] =\frac{e_{0}}{\mathcal{V}m} \mathrm{Tr} \left[\rho^g (\mathbf{p}-e_{0}\mathbf{A}^{g})\right].
\label{Jpure}
\end{equation}
It is important to stress here, that the source of any secondary radiation (such as high harmonics) is not the canonical, but the kinetic momentum $\mathbf{p}%
_{kin},$ due to its direct connection to the velocity operator \cite{CT04}.

Let us look at the expectation value above in the two gauges we consider here. Similarly to the more detailed calculations presented in the Appendix, in
the velocity gauge we obtain
\begin{align}
\mathbf{J}&=\frac{e_{0}}{\mathcal{V}m}\sum\limits_{n\mathbf{k}}|c_{n\mathbf{k}}(t)|^{2}\left\{
\hbar\mathbf{k-}e_{0}\mathbf{A(}t\mathbf{)}\right\} \label{Jv} \\
&-i\hbar\frac{e_{0}}%
{\mathcal{V}m}\sum\limits_{n,n^{\prime}}\sum\limits_{\mathbf{k}}c_{n\mathbf{k}}^{\ast
}(t)c_{n^{\prime}\mathbf{k}}(t)\int_{\Omega}u_{n,\mathbf{k}}^{\ast}%
(\mathbf{r})\nabla u_{n^{\prime},\mathbf{k}}(\mathbf{r})d^{3}\mathbf{r}.
\nonumber
\end{align}
In the length gauge $\mathbf{A}^{l}\mathbf{=}0$, and we have therefore:
\begin{align}
\mathbf{J} &  \mathbf{=} \frac{e_{0}}{\mathcal{V}m} \mathrm{Tr}\left[ \rho^g \mathbf{p}%
_{kin}\right] =\frac{e_{0}}{\mathcal{V}m}\sum\limits_{n\mathbf{k}}\hbar\mathbf{k}|b_{n\mathbf{k}%
}(t)|^{2} \label{Jl}\\
&-i\hbar\frac{e_{0}}{\mathcal{V}m}\sum\limits_{n,n^{\prime},\mathbf{k}}b_{n\mathbf{k}}^{\ast}(t)b_{n^{\prime}\mathbf{k}}
(t)\int_{\Omega}u_{n,\mathbf{k}}^{\ast}(\mathbf{r})\nabla u_{n^{\prime
},\mathbf{k}}(\mathbf{r})d^{3}\mathbf{r}.\nonumber
\end{align}

The results calculated in any of the gauges must be identical,
but this equality is valid only for the \emph{whole expressions in Eqs.~(\ref{Jv}) and (\ref{Jl}),} and \emph{by no means for the
individual terms on the right hand sides. }

\bigskip

\emph{Gauge dependence } It is tempting to regroup the terms in Eqs.~(\ref{Jv}) and (\ref{Jl}) and interpret the macroscopic, "current-like" part of the source of the HHG radiation
\begin{equation}
\mathbf{j}^{g}=\frac{e_0}{\mathcal{V}m}\sum_{n \mathbf{k}} \left|c_{n\mathbf{k}}^{g}(t)\right|^2 \langle n,\mathbf{k}|\mathbf{p}_{kin}|n,\mathbf{k}\rangle,
\label{diagHHG}
\end{equation}
as the intraband contribution, while the microscopic, "polarization-like" part
\begin{equation}
\dot{\mathbf{P}}^{g}= \frac{e_0}{\mathcal{V}m}\sum_{n\neq n' \mathbf{k}} \left[c_{n\mathbf{k}}^{g}(t)\right]^*c_{n'\mathbf{k}}^{g}(t) \langle n,\mathbf{k}|\mathbf{p}_{kin}|n',\mathbf{k}\rangle
\label{offdiagHHG}
\end{equation}
as the interband contribution. Indeed, $\langle n,\mathbf{k}|\mathbf{p}_{kin}|n,\mathbf{k}\rangle/m$ is the band velocity in the $n$th band at point $\mathbf{k},$ thus $\mathbf{j}^{(g)}$ means the sum of all these velocities weighted by the corresponding populations. For comparison with textbook methods \cite{HK04}, let us note that $\langle n,\mathbf{k}|\mathbf{p}|n,\mathbf{k}\rangle$ can also be calculated as $m/\hbar \nabla_{\mathbf{k}} E_{n}(\mathbf{k}),$ but Eqs.~(\ref{Jv}) and (\ref{Jl}) treat inter- and intraband currents on an equal footing. Clearly, $\mathbf{J}=\mathbf{j}^{g}+\dot{\mathbf{P}}^{g},$ and %-- according to Eq.~(\ref{JJ}) --
 it is gauge independent, but it does not necessarily hold for $\mathbf{j}^{g}$ and $\dot{\mathbf{P}}^{g}$ separately.

Let us investigate the relation (\ref{coefflv}) between the expansion coefficients $c_{n\mathbf{k}}(t)$ and $b_{n\mathbf{k}}(t).$  Assume that the dynamics is solved in the velocity gauge [i.e., $c_{n\mathbf{k}}(t)$ are known functions of time], and let us transform the result to length gauge. We obtain:
\begin{align}
\nonumber
&b_{n'\mathbf{k}'}(t)=\sum\limits_{n\mathbf{k}%
}\left\langle n^{\prime}\mathbf{k}^{\prime}\right\vert \exp\left[
-ie_{0}\mathbf{A(}t\mathbf{)}\mathbf{r/}\hbar\right]  |n,\mathbf{k}\rangle
c_{n\mathbf{k}}(t) \\
\nonumber
&=\sum\limits_{n\mathbf{k}} \int_{\mathcal{V}} e^{i\mathbf{r}(\mathbf{k}-e\mathbf{A}(t)/\hbar+\mathbf{k}')} u_{n',\mathbf{k}'}^*(\mathbf{r})   u_{n,\mathbf{k}}(\mathbf{r}) d^3\mathbf{r}  \ c_{n\mathbf{k}}(t)\\
\label{bnk}
&= \sum\limits_{n} \int_\Omega u_{n',\mathbf{k}'}^*(\mathbf{r})   u_{n,\mathbf{k}(t)}(\mathbf{r}) d^3\mathbf{r} \ c_{n\mathbf{k}(t)}(t),
\end{align}
where $\mathbf{k}(t)=e\mathbf{A}(t)/\hbar-\mathbf{k}'.$ The first consequence of this result is that whenever the velocity gauge density operator is diagonal in $\mathbf{k},$ [which is the case during the whole time evolution for the initial conditions given by Eq.~(\ref{rho0})], the same holds also for the length gauge, but with $\mathbf{k}$ values being  shifted by $e\mathbf{A}(t)/\hbar$. [This is in agreement with the acceleration theorem (\ref{accelerationeq}).] On the other hand, since the lattice-periodic functions $u_{n,\mathbf{k}}$ are orthogonal only for the same index $\mathbf{k},$ the integral in the last line of Eq.~(\ref{bnk}) is not proportional to $\delta_{nn'}.$

This effect is closely related to the fact that the separation of $\mathbf{J}$ as the sum of intraband ($\mathbf{j}$) and interband ($\dot{\mathbf{P}}$) components is not unambiguous. E.g., for $\mathbf{j}$ we have
\begin{equation}
\mathbf{j}^{v}=\frac{e_0}{\mathcal{V}m}\sum_{n \mathbf{k}} \left|c_{n\mathbf{k}}(t)\right|^2 \langle n,\mathbf{k}|\mathbf{p}|n,\mathbf{k}\rangle -\frac{\mathcal{N}e_0^2}{\mathcal{V}m} \mathbf{A},
\label{diagHHGv}
\end{equation}
where $\mathcal{N}=\mathrm{Tr}\rho.$ On the other hand,
\begin{equation}
\mathbf{j}^{l}=\frac{e_0}{\mathcal{V}m}\sum_{n \mathbf{k}} \left|b_{n\mathbf{k}}(t)\right|^2 \langle n,\mathbf{k}|\mathbf{p}|n,\mathbf{k}\rangle.
\label{diagHHGl}
\end{equation}
The last term in Eq.~(\ref{diagHHGv}) is determined solely by the time dependence of the vector potential, while the coefficients $c_{n\mathbf{k}}$ and $b_{n\mathbf{k}}$ that mean the difference between the sums in Eqs.~(\ref{diagHHGv}) and (\ref{diagHHGl}) are related via integrals that are eventually determined by the structure of the material. This means that even if for a certain material and time dependent vector potential $\mathbf{j}^{l}(t)$ and $\mathbf{j}^{v}(t)$ may happen to be identical, there is no physical reason why this should hold e.g., for a different target material. In other words, the inter- and intraband  contributions that constitute the gauge independent source term $\mathbf{J},$ cannot be determined without referring to a particular gauge.

\bigskip

\emph{Numerical example }
\label{numericalsec}
\begin{figure}[tbh]
\includegraphics[width=8.5cm]{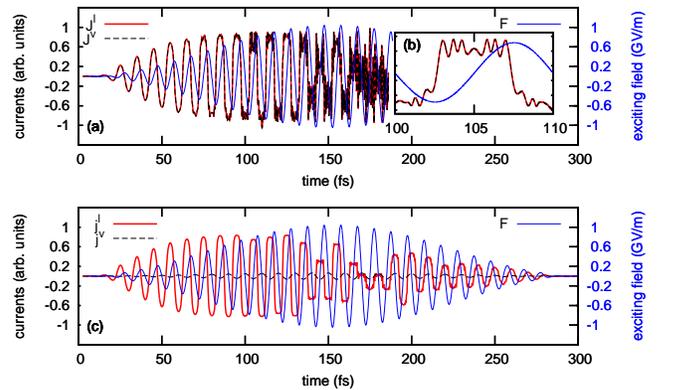}
\caption{Panel (a): The total current $J$ (in both of the considered gauges) and the exciting field as a function of time. Panel (b) zooms on a certain peak. Panel (c): the intraband current $j.$ For the sake of simplicity, the initial density operator is $\rho_0=\vert n_0,0\rangle\langle n_0,0\vert.$ The central wavelength of the excitation pulse is $\lambda=3 \mu$m.}%
\label{2bandHHGfig}
\end{figure}
As an illustration, we calculate the current generated by a short electromagnetic pulse in a one-dimensional model crystal, where the exciting field is polarized along the periodic chain. We use a model potential that produces a direct bandgap of 3.2 eV, similarly to the case of ZnO \cite{GDiCSADiMR11}.
Having determined the Bloch-sates, the time dependence of $j^g$ and $\dot{P}^g$ [the one-dimensional versions of the quantities given by Eqs.~(\ref{diagHHG}) and (\ref{offdiagHHG}), respectively] can be calculated, as well as their sum, $J^g.$ [The superscript reminds us the fact that various approximations may result in the (incorrect) gauge dependence of $J.$]

As we can see in the bottom panel of Fig.~\ref{2bandHHGfig}, the time dependence of $j^v$ and $j^l$ are clearly different.  The total currents $J^v$ and $J^l$ are not exactly the same either, but for the 8-band model presented in Fig.~\ref{2bandHHGfig}, the simple one-dimensional numerical model supports our earlier findings within reasonable precision.

\emph{Discussion}
An overview of the results above shows that the main reason why one cannot make a gauge-independent distinction between intra- and interband dynamics is related to the fact that states corresponding to different gauges are not identical. As an analogue, let us recall that a similar statement holds also for a single atom in an electromagnetic field. Even in dipole approximation, depending on the choice of the gauge, the states may "incorporate some interaction with the field" \cite{BMB05}. However, for optical frequencies and laser field strengths around 1 GV/m, the phase factor appearing in $\exp(-ie\mathbf{A}\mathbf{r}/\hbar)$ is almost constant over the size of an atom, its change is around $\pi/100,$ thus the effect is weak. In the case of solids, on the other hand, the interaction region is much larger, and the effect cannot be neglected.

Finally, let us emphasize, that all the considerations above are relevant \emph{during} the excitation. When the exciting pulse is over (i.e., $\mathbf{A}=0$), the separation $\mathbf{J}=\mathbf{j}+\dot{\mathbf{P}}$ is clearly unambiguous, and e.g.,~$\mathbf{j}$ -- as a macroscopic current -- can be measured.

\bigskip

\emph{Summary}
We investigated the process of high-order harmonic generation in bulk solids. Working in the single electron picture, light-matter interaction were considered in two widely used electromagnetic gauges. Inter- and intraband contributions to the whole current were treated in a unified way. Both analytic and numerical results show that the interpretation of the process in terms of inter- and intraband currents has to performed with care, since these contributions do depend on the choice of the electromagnetic gauge.

\section*{Acknowledgments}
The author thanks M.~G.~Benedict, S.~Varr\'o, K.~Varj\'u and A. Czirj\'ak for useful discussions. Partial support by the ELI-ALPS project is acknowledged. The ELI-ALPS project (GOP-1.1.1-12/B-2012-000, GINOP-2.3.6-15-2015-00001) is supported by the European Union and co-financed by the European Regional Development Fund. The work was also supported by
the Hungarian National Research, Development and Innovation
Office under Contract No.~124750.

\section{Appendix}
\subsection{Bloch-basis matrix elements}
For practical calculations, the matrix elements of the Hamiltonian in the Bloch-state basis are very useful. Remaining in the dipole approximation, where the spatial dependence of
the field amplitude can be neglected, we obtain:
\begin{align}
&\langle n,\mathbf{k}|\mathbf{p}|n^{\prime},\mathbf{k}^{\prime}\rangle =\nonumber \\
&=-i\hbar \int_{\mathcal{V}}\Psi_{n,\mathbf{k}}^{\ast}(\mathbf{r})\nabla
\Psi_{n^{\prime},\mathbf{k^{\prime}}}(\mathbf{r})d^{3}\mathbf{r}
\nonumber \\
&=-\frac{i\hbar}{\mathcal{V}}
\int_{\mathcal{V}}e^{i\mathbf{r}(\mathbf{k^{\prime}-k})}u_{n,\mathbf{k}}%
^{\ast}(\mathbf{r})(i\mathbf{k}^{\prime}+\nabla)u_{n^{\prime},\mathbf{k}^{\prime}%
}(\mathbf{r})d^{3}\mathbf{r}\nonumber\\
&=-\frac{i\hbar}{\mathcal{V}}\sum_{\mathbf{R}}e^{i\mathbf{R}(\mathbf{k^{\prime}-k})}\int
_{\Omega}u_{n,\mathbf{k}}^{\ast}(\mathbf{r})(i\mathbf{k}^{\prime}%
+\nabla)u_{n^{\prime},\mathbf{k}^{\prime}}(\mathbf{r})d^{3}\mathbf{r}%
\nonumber\\
&  =\delta_{\mathbf{k}\mathbf{k}^{\prime}}\left[  \hbar\mathbf{k}%
\delta_{nn^{\prime}}-i\hbar\int_{\Omega}u_{n,\mathbf{k}}^{\ast}(\mathbf{r}%
)\nabla u_{n^{\prime},\mathbf{k}^{\prime}}(\mathbf{r})d^{3}\mathbf{r}\right],
 \label{pmatrixl}%
\end{align}
where $\mathbf{R}$ denotes lattice vectors.
In view of this, the matrix elements of $H_{ext}$ in velocity gauge [recall that $\mathbf{A(}t\mathbf{)\equiv A}^{v}%
\mathbf{(}t\mathbf{)}$] are given by:
\begin{align}
&\langle n,\mathbf{k}|\frac{1}{2m}\left[  -2e_{0}\mathbf{p}\mathbf{A}\mathbf{(}t\mathbf{)}+e_{0}^{2}A^{2}(t)\right]  |n^{\prime}%
,\mathbf{k}^{\prime\text{ }}\rangle=
\nonumber \\
&=-\frac{e_{0}}{m}\langle n,\mathbf{k}%
|\mathbf{p}|n^{\prime},\mathbf{k}^{\prime}\rangle\mathbf{A(}t\mathbf{)+}%
\frac{1}{2m}e_{0}^{2}A^{2}(t)\delta_{\mathbf{k}\mathbf{k}^{\prime}}%
\delta_{nn^{\prime}}\nonumber\\
&=\delta_{\mathbf{k}\mathbf{k}^{\prime}}\left[  -i\hbar\int_{\Omega
}u_{n,\mathbf{k}}^{\ast}(\mathbf{r})\nabla u_{n^{\prime},\mathbf{k}^{\prime}%
}(\mathbf{r})d^{3}\mathbf{r}\right]  \mathbf{A(}t\mathbf{)}
\nonumber \\
&+\left[
\hbar\mathbf{k\mathbf{A(}}t\mathbf{\mathbf{)}+}\frac{1}{2m}e_{0}^{2}%
A^{2}(t)\right]  \delta_{\mathbf{k}\mathbf{k}^{\prime}}\delta_{nn^{\prime}}.
\label{Hmatrixelements}
\end{align}
Note that the reason why these equations have a relatively simple form is the compatibility of the momentum operator with the translational symmetry of the crystal.
On the other hand, the matrix elements of $H_{ext}^{l}(t)$ as
given by $-e_0\langle n,\mathbf{k}|\mathbf{r}|n^{\prime},\mathbf{k}^{\prime}\rangle\mathbf{F}(t)$
cannot be simplified as above, because the operator $\mathbf{r}$ is not lattice-periodic.

\subsection{Numerical details}
In order to be able to calculate the matrix elements as discussed above, we need the actual Bloch-states, that is, in a discretized picture, the complex values $u_{nk}(x_m),$ where $x_m$ is in the unit cell ($-a/2<x_m<a/2$). For the sake of definiteness, the lattice constant $a$ is chosen to be 0.5 nm. Once the potential is specified, these eigenfunctions can be obtained using
\begin{equation}
\left[\frac{\hbar^2}{2m}\left(-i\frac{\partial}{\partial x} +k\right)^2 + V(x)\right]u_{n,\mathbf{k}}(\mathbf{r})=E_n(k)u_{n,\mathbf{k}}(\mathbf{r}).
\label{ueigen}
\end{equation}
As numerical examples, we investigated two model potentials.  $V_1$ contains two localized attracting centers:  $V_1(x)[eV]=-25\cos^2\left[\pi(x-x_1)/(15 a)\right]$ $-25\cos^2\left[\pi(x-x_2)/(15 a)\right]$, where $x_1/a=-0.2$ and $x_2/a=0.107,$ and the $\cos^2$ functions are zero unless their argument is in the interval $[-\pi/2,\pi/2]$. The second model potential is a modification of the one that has been used in Ref.~\cite{HIY15}: $V_2[eV]=−25[1 + \tanh(x + x_0)][1 + \tanh(−x +x_0)],$ where $x_0=0.2475 a_0$ ($a_0$ denotes the Bohr radius). There is an important point these potentials have in common: both produce a bandgap of 3.2 eV, which is in agreement with the material properties of ZnO. Clearly, apart from the bandgap, the two potentials correspond to different band schemes, but the qualitative results related to the question of gauge invariance are the same in both cases. For the sake of definiteness, we used $V_1$ for obtaining the data presented in Fig.~1.

Having $N$ grid points $x_n$ in the 1D unit cell, Eq.~(\ref{ueigen}) with periodic boundary conditions provide a set of eigenfunctions $\left\{u_{n,k}\right\}_1^N$ that forms a basis for each value of $k.$ (That is, any finite-valued periodic function defined on the grid can be written as a linear combination of these eigenfunctions.) It is convenient to use discrete $k$ indices as well, $k_m=-\pi/a+(m-1)\Delta k, \ m=1,\ldots, M$ and $(M-1)\Delta k=2\pi/a.$ In this way, we obtain $N\times M$ functions $u_{n,k},$ and -- besides the matrix elements (\ref{Hmatrixelements}) -- we can calculate their overlaps
\begin{equation}
S_{nn'}^{kk'}=\int_{-a/2}^{a/2} u_{n,k}^{*}(x) u_{n',k'}(x) dx
\label{overlap}
\end{equation}
numerically. Note that -- by construction -- we have $S_{nn'}^{kk}=\delta_{nn'}.$ With sufficient number of real-space discretization points, the integrals (\ref{Hmatrixelements}) and (\ref{overlap}) converge for the lowest lying energy eigenstates. (According to our experience, it is sufficient for both $N,M$ to have the order of magnitude of a few hundred.)

In velocity gauge, the index $k$ of the Bloch-states does not change during the time evolution, and by inspecting the populations we can determine how many bands play observable role in the time evolution. Using the appropriate (converged) matrix elements, we can follow the time evolution, and transform the result to length gauge at appropriate time instants. More precisely, for a given initial index $k_0,$ whenever $\mathbf{k}(t)=e\mathbf{A}(t)/\hbar-\mathbf{k}_0$ equals one of the k-space grid points $k_m$, we can use the overlap matrix to calculate the transformation. Since these time instants $t_m$ are the same for any initial index $k_0,$ the time evolution of the complete density matrix can be calculated and compared in both gauges at the discrete $t_m$ values. In this way, discretization in k-space affects only the time resolution of the results: at $t_m,$ the calculated physical quantities are numerically exact in the sense that the only approximation is real space discretization.

In our actual calculations, the exciting laser field is assumed to be polarized along the $x$ direction, and the time dependence of the only nonzero component of the vector potential is given by
\begin{equation}
A(t)=A_0 \sin^2\left(\frac{\pi t}{T}\right) \cos(\omega t),
\label{pulse}
\end{equation}
provided $t\in[0,T],$ and zero otherwise. Similarly to Ref.~\cite{GDiCSADiMR11}, we consider many-cycle, mid-infrared excitation. That is, $\lambda=3 \mu$m and $T=300$ fs in the calculations. The amplitude is chosen to correspond to a peak laser field strength of 1.0 GV/m.

Finally, note that the overlap matrix (\ref{overlap}) is clearly unitary, provided all the bands are taken into account. However, when we do not use all the bands (either below or above the highest valence band), the appropriate projection of $S$ may not be exactly unitary, leading to the fact that the norm (trace of the density matrix) is not the same in the two gauges. In other words, besides the usual criteria a numerical method has to satisfy (convergence, stability, etc.), the requirement of gauge invariance sets a new one. Working in Bloch-state basis, this simply means that a large enough number of bands has to be taken into account.

\end{document}